\documentstyle[preprint,aps,psfig]{revtex}

\begin{document}
\draft
\title{Capture on High Curvature Region: 
Aggregation of Colloidal Particle Bound to Giant Phospholipid Vesicles}
\author{Quan-Hui Liu$^{1,2}$\cite{l}, Ji-Xing Liu${^1}$, Ou-Yang Zhong-Can${^1}$}
\address{$^{1}$Institute of Theoretical Physics, Academia Sinica, P.O. Box 2735, Beijing, 100080, China}
\address{$^{2}$ Department of Physics, Hunan University, Changsha, 410082, China}
\date{\today}
\maketitle

\begin{abstract}
A very recent observation on the membrane mediated attraction and
ordered aggregation of colloidal particles bound to giant phospholipid
vesicles (I. Koltover, J. O. R\"{a}dler, C. R. Safinya, Phys.\ Rev.\ Lett. 
{\bf 82}, 1991(1999)) is investigated theoretically within the frame of
Helfrich curvature
elasticity theory of lipid bilayer fluid membrane. Since the concave or waist regions of the
vesicle possess the highest local bending energy density, the aggregation of colloidal 
beads on these places can reduce the elastic energy in maximum. Our calculation 
shows that a bead in the concave region lowers its energy $\sim 20 k_B T$. For an 
axisymmetrical dumbbell vesicle, the local curvature energy density along the waist 
is equally of maximum, the beads can thus be distributed freely with varying separation distance. 
\end{abstract}


\preprint{Submitted to PRL}


Recently, much attention focused on the experimental results carried out by a
group in University of California (GUOC) on the interaction between DNA,
colloidal particles and lipid membrane\cite{saf1,saf2,saf3,sci,exp1,exp2}.
These experimental advances are helpful to understand the structure of
bioassembles, the possible vesicle shape transformations, the capture
function and the transport processes of the membrane, etc.; and all of these
have suggested their potential use for industrial as well as medical
applications\cite{sci}. In 1998, even two independent theoretical groups
gave the same theory to explain the structure of DNA-cationic lipid
complexes observed by the GUOC\cite{exp1,exp2}. In March of 1999, the
GUOC reported a light microscopy study of the interactions between
colloidal beads either chemically bound or physisorbed  onto 
flexible giant vesicle fluid membranes\cite{saf3}. But the chemically bound
cases were a continuous work of their previous one\cite{saf1,saf2}. 
Their new experiments 
were mainly on the physisorption between the colloidal beads and membranes. 
They observed two essentially new phenomena via two types of different experiments. 

As GUOC reported, the initial bare unilamellar vesicles of sphere topology were prepared from phospholipids POPC 
(1-Palmytoyl-2-Oleoyl Phosphocholine) or DMPC (Dimiristoyl Phosphocholine) and 
$ 0.5 \% $ biotin-X DHPE (molecular probes) following a standard protocol\cite{saf3}, 
and the typical vesicle has a radius $ R\sim 10\mu m $. The colloidal beads were chemically grafted with treptavidin $--$ a water
soluble protein which specifically binds to biotin; and the typical bead has a radius $a\sim 0.45\mu m$. 
After putting the beads to the surfaces of vesicles, treptavidin 
beads were linked to giant vesicle doped with biotinilated lipids by mixing dilute suspensions of each component.
First type of their experiment is on the shapes of 
equilibrium bead-membrane hybrid vesicles. They found that these vesicles clearly revealed 
concave regions where the beads locate. For further demonstration of this fact, 
they had captured vesicles using suction glass pipettes then released them, and 
found that the vesicles reverted to the nearly original shapes with the bead 
centered at concave regions again. We noticed that when the bead-membrane hybrid vesicles are either in free
states or in head of the suction glass pipettes, the bead pinched in membrane and the degree of
pinch$--$which is balanced by the adhesion force$--$does not appear significant change. 
When there are
many beads on vesicles, these beads either segregated condensed phases of beads located in 
concave regions of membranes, or firmly pinned to the waist of a nearly axisymmetrical dumbbell
shape vesicle along the meridional direction but distributed freely with varying separation 
distance along the azimuthal direction. The second type of their experiment is on
the movement of the two beads having the same radius on a quasi-spherical vesicle. At first, these two
beads walk randomly in the surface of the vesicle. If their center-center distance was within four times of their radius,
an attraction drives them together. These two beads finally formed 
a concave region and remained there from then on. 

From their experiments, two conclusions seem to be definite. 
First is that there may be no attraction even on the membrane otherwise the 
beads in the waist of an axisymmetrical dumbbell
shape vesicle could be condensed. Of course, there is no evidence to 
show that there are observable repulsion between them. Second, whenever the vesicles adhered by beads 
are in free states or are captured by suction glass pipettes, and whenever the beads are randomly walking 
on of the membranes or trapped in concave region, there is no observable completely enclosed or laid bare 
phenomena for the pinched beads. It means that the adhesion energy of each bead is nearly the same as long as it is on the 
membrane. We have noted that the GUOC has proposed a theoretical explanation for such phenomena based on a known theory developed by Dan
\cite{dan}, and indeed offered many useful theoretical information. But their explanation is not 
sufficient enough to put all of their results into a theory coherently. For example,
from the Dan's theory, the membrane-induced attractions are caused by the adsorbing
DNA perturbing the equilibrium packing of the lipids, and the range of interaction is  
typically of order $n m$. But here the distance of two beads starts attracting is $1.8 \mu m$ and 
each bead at least four orders heavier than the DNA molecula, and there is no attraction between beads in 
the waist line of an axisymmetrical dumbbell. Whether we can apply the principle of Dan's theory to our situation
seems to be a problem. Moreover, the concave regions seem not to be represented by 
catenoid shapes as did in\cite{saf3}, because the mean curvatures and therefore the curvature energies of such shapes are zero,
 and any further deformations of the membranes will increase the elastic energy. 
Thus the localization of beads on concave regions is energetically
unfavored. In this Letter, we will present an improved theory to account for all these phenomena. 
In our approach, the concave or waist regions of the
vesicle possess the highest local bending energy density, the aggregation of colloidal 
beads on these places can reduce the elastic energy in maximum. An explicit calculation will
show that a bead in the concave region lowers its energy $\sim 20 k_B T$ where $k_B$ is the Boltzmann constant and $T$ 
the temperature. For an 
axisymmetrical dumbbell vesicle, the local curvature energy density along the waist 
is equally of maximum, the beads can thus be distributed freely with varying separation distance. 

First of all, we take for grant, as experiments implied (a quantitative calculation will be given latter),  
that the area of pinching membranes around each bead of the same radius is fixed and 
the adhesion energy for each bead then is regarded as the same as long 
as the beads are on the membrane. The stable shapes of bead-membrane hybrids are of course energy minimum states, and 
the two energy degenerate shapes of 
sphere topology vesicle are axisymmetrical biconcave shape and the axisymmetrical dumbbell shape from the 
instability analysis of a sphere shape vesicle\cite{oy}. So, the  
formations of concave regions or waist for sphere can be taken place either before the beads were introduced, or 
while the bead-membrane hybrids look for the energy minimum states. According to experimental facts\cite{saf3}, 
if there only one bead, the bead will locate on the concave region; if there are many beads on it, there must be a bead at the center of the concave region first for that a 
bead-membrane hybrid vesicle after all develops a concave region to trap these beads. The main text of this Letter is devoted
to show why a bead favors the concave regions and how two beads attract in detail.
For our purpose, we will use an exact solution representing a concave shape in
the fluid membrane to show that the free energy density per area in the vesicle is not uniformly distributed;  
and seriously deformed areas, concave and waist regions for example, have the much higher energy density 
than elsewhere. Location of the beads in these places can make the whole energy lower down.
The membrane mediated attraction is thus induced by the free energy difference. 

In physics, equilibrium shapes of phospholipid vesicles are assumed to
correspond to the minimum of the elastic energy of the closed bilayer
membrane consisting of the amphiphilic molecula. The energy functional or the 
total free energy $F$ of a vesicle is given by Helfrich spontaneous curvature theory as\cite{hel}: 
\begin{equation}
F={\frac{1}{2}}k_c\int (c_1+c_2-c_0)^2dA+\delta p\int dV+\lambda \int dA,
\label{e1}
\end{equation}
where $dA$ and $dV$ are the surface area and the volume elements for the
vesicle, respectively, $k_c \sim 30 k_B T$ is an elastic modulus\cite{saf3,hel}, $c_1$ and $c_2$ are the
two principal curvatures of the surface, and $c_0$ is the spontaneous
curvature to describe the possible asymmetry in both sides of the bilayer membrane. The
Lagrangian multipliers $\delta p$ and $\lambda $ take account of the
constraints of constant volume and area, which can be physically understood
as the osmotic pressure between the ambient and the internal environments,
and the surface tension, respectively. In GUOC experiment, the tension energy 
$\lambda \int dA$ and the pressure energy $\delta p\int dV$ appear not to change significantly,
then we will simply discard these two constant terms hereinafter. Here 
we have not included the interaction between the beads and membranes, but
will treat them as perturbations. 

Variational calculus of the elastic energy Eq.(\ref{e1}) gives the equilibrium shape
equation\cite{oy}: 
\begin{equation}
(2H+c_0)(2H^2-2K-c_{0}H)+ 2 {\bigtriangledown}^2 H=0,  
\label{e2}
\end{equation}
where ${\bigtriangledown}^2={\frac{1 }{\sqrt{g}}}{\partial}_{i}(g^{ij}\sqrt{g
}{\partial}_{j})$ is the Laplace-Beltrami operator, $g$ is the determinant
of the metric $g_{ij}$ and $g^{ij}=(g_{ij})^{-1}$, $K=c_1 c_2$ is the
Gaussian curvature and $H=-(1/2)(c_1+c_2)$ is the mean curvature. Assuming
that the shape has axisymmetry, the general shape equation Eq.(2) becomes a
third order nonlinear differential equation\cite{oy}: 
\begin{eqnarray}
\cos^{3}\psi(\frac{d^{3}\psi}{dr^{3}}) & = & 4\sin\psi\cos^{2}\psi( \frac{d^{2}\psi}{dr^{2}})(\frac{d\psi}{dr}) - \cos\psi(\sin^{2}\psi- \frac{1}{2}
\cos^{2}\psi )(\frac{d\psi}{dr})^{3}  \nonumber \\
& + & \frac{7\sin\psi\cos^{2}\psi}{2r}(\frac{d\psi}{dr})^{2} - \frac{2\cos^{3}\psi}{r}(\frac{d^{2}\psi}{dr^{2}})  \nonumber \\
& + & [\frac{c_{o}^{2}}{2} - \frac{2c_{o}\sin\psi}{r} - 
\frac{\sin^{2}\psi-2\cos^{2}\psi}{2r^{2}}]\cos\psi(\frac{d\psi}{dr}) 
\nonumber \\
& + & [ \frac{c_{o}^{2}\sin\psi}{2r} - \frac{\sin^{3}\psi+2\sin\psi\cos^{2}\psi} {2r^{3}}].
\label{oyhu}
\end{eqnarray}
where $r$ is the distance from the symmetric $z$ axis of rotation, $\psi(r)$
is the angle made by the surface tangent and the $r$ axis.
The positive direction of the angle is that the angle measured clockwise
from $r$ axis. It is contrary to the usual mathematical convention; so, the
mean curvature $H$ is $-1/2(\sin\psi/r+d\sin\psi/dr)$, in which $c_1=\sin\psi/r $ denotes the principal curvature along the parallels of
latitude, and $c_2=d\sin\psi/dr$ denotes that along those of meridians. To note
that different papers may use different sign conventions, and the sign convention 
used in this Letter is self-consistent. It
is worthy of mentioning that the spontaneous curvature $c_0$ carries a sign.
When the normal of a surface change its direction, $c_1, c_2$ and $c_0$ must
change their signs simultaneously. Keeping the directions of $r$ and $z$ as
the usual, we have therefore: 
\begin{equation}
\left\{ 
\begin{array}{l}
dz/dr=-\tan\psi(r) \nonumber \\ 
z(r)-z(0) =-\int ^{r}_0 \tan\psi(r) d r.
\end{array}
\right.
\end{equation}
We see that the sphere with parameterization $\sin \psi (r)=r/R_0$, where $R_0>0$ 
is the radius of the sphere, is always a solution of Eqs. (\ref{e2},\ref{oyhu}). It is 
easily to verify that Eqs. (\ref{e1},\ref{e2}) have the following solutions\cite{oy93}  
\begin{equation}
\sin \psi =r+c_0r \ln r,  \label{sin}
\end{equation}
where we take $r$ is measured with $10\mu m$ as the length unit and so for the value of spontaneous curvature $c_0$. 
When $c_0=0$, this solution gives nothing but sphere of radius $10\mu m$. When $c_0$ is a
small quantity, positive $c_0$ leads to oblate ellipsoid and negative $c_0$
leads to prolate ellipsoid respectively\cite{liu}. Increasing $c_0$ from zero, we can have 
a great variety of the concave regions from shallow to deep. 
The elastic energy density, the energy per area, from Eqs.(\ref{e1},\ref{sin}) is:  
\begin{equation}
dF/dA=2 k_c (1+c_0\ln r)^2 .
\label{ener}
\end{equation}

To obtain a typical concave region, an average shape of concave regions trapping beads in the Figs.1, 2, 5, of GUOC report\cite{saf3} is
used for fitting, and such a concave region is shown in Fig.1 by the upper half of a sphere of radius $10 \mu m$. 
Its parameterization is $\sin \psi =r+2.1r \ln r$, and we plot the energy density by thin line in Fig.2. 
The elastic energy in the whole concave region, i.e. in the upper half piece of the surface of quasi-sphere shape, can be 
easily carried out to be $280.61k_BT$. 
The contact area $\sigma $ between each bead and membrane are reasonably
$\sigma =\pi r_0^2=\pi (0.3)^2 (\mu m)^2$ as shown in \cite{saf3} with $r_0=0.3\mu m$ being the projected radius\cite{saf3},
and the adhesion energy is comparable with 
elastic energy in this contact area, which is $2k_c\sigma /a^2\sim 100k_BT$\cite{saf3}. This is 
convincing because it guarantees that the beads can never escape from the vesicle due to the 
thermal agitation. Since the adhesion energy is independent from the place where the beads locate, 
the more serious deformed the original membrane is, the more bending energy is released from the whole free energy. 
The original bending energy in the contact area of the membrane then plays the role of trapping energy. Since the 
beads constrained in the membrane, it has only two degrees of freedom. The kinetic energy of each bead is therefore 
$2 (1/2 k_BT)=k_BT$. If the trapping energy is less than this energy, the beads reveal Brownian walk motion. 
The trapping energy in the center of the concave region is $ \Delta F(0) $:
\begin{eqnarray}
\Delta F(0)&&=2 k_c \int _{\sigma}(1+2.1lnr)^2 dA \nonumber \\
       &&=2 k_c \int _{0}^{0.03}(1+ 2.1lnr)^2/\sqrt{1-(r +2.1 r ln r)} 2 \pi r dr \nonumber \\
       &&=19.8 k_B T\sim 20 k_BT.
\label{energy1}
\end{eqnarray}
The trapping energies in range $ r\geq 2\mu m $ are given by
$ \Delta F(r) = 2 k_c (1+2.1lnr)^2 \sigma =  2 \pi(0.3)^2 k_c (1+2.1lnr)^2 < 0.96 k_B T$, as shown by the bold line of the Fig. 2.  
Therefore in these interval, the Brownian motion would dominate the bead's trajectory.
When $r=4*0.45\mu m=1.8\mu m$, $ \Delta F(1.8\mu m)=1.15 k_B T $,  the trapping energy is comparable with the thermal fluctuation.   
Once the bead is within  $r\leq 1.8\mu m$, the trapping energy overwhelms the thermal fluctuation and 
would drive the bead move to the center. Once a bead moves to the center, its Brownian motion is energetically depressed 
and the bead can be regarded as trapped. This gives a reasonable explanation to the primary experimantal result as the tendency 
of a bead to locate on a concave region as observed in \cite{saf3}.

For further demonstration the degree of agreement of our theory with the experimental date, we 
estimate the magnitude of relative velocity $v$ of the motion of two beads from relative 
position $r=1.8\mu m$ to $r=0.9\mu m$ with one bead already locating at the center of the concave region as shown in Fig. 1.  
We have approximately $1/2 m v^2\sim 1k_BT-10 k_BT$, i.e., we have $v \sim v_0-3 v_0$ with $v_0=\sqrt{2 k_BT/m}$. By comparing these date with the experimental ones,
we see that agreement is twofold. First it means that in whole range $r<1.8\mu m$, 
there is an increase of velocity from $v_0$ to $3 v_0$; it is true from the observation as shown in Fig.3(b)\cite{saf3}. 
Second, the average of interaction strength in the range $r=1.8\mu m$ to $r=0.9\mu m$ is $5 k_BT$, which can be
regarded as the quantitative statement of ``several $k_B T$''\cite{saf3}. 

For checking the above theoretical consistence of our approach, we return to verify the basic assumption about 
the constancy of the contact area $\sigma=\pi (0.3)^2(\mu m)^2$, i.e., the constancy of the adhesion 
energy. Since the elastic energy has the scaler invariance, all sphere shapes have the same energy $8 \pi k_c\sim750 k_BT$,
independent from the radii\cite{serf}. Therefore, if a small change of the contact area as $\sigma=\pi (0.3\pm 0.1)^2(\mu m)^2$
is allowed, it would accompany the large energy change $\Delta F=37.21k_BT-148.95 k_B T$. So large energy change on 
the membrane must have the observable transport effect of membrane materials, but  
no such processes were reported in the experiment\cite{saf3}. We would also like to compare the 
elastic energy of a perfect sphere shape vesicle with it having a concave region in its half piece as shown in Fig. 1, 
the elastic energy reduces an amount $376.46k_BT-280.61k_BT=95.85 k_BT $. 
Therefore the formation of the concave regions is also energetically favored.

Till now we have developed a consistent theoretical approach to explain the primary experimental results of GUOC\cite{saf3}:
The tendency of a bead to be captured on the center of a concave region (the region of high curvature region), 
and the membrane mediated attraction due to the elastic energy difference of the membrane. Such principle can be
used to explain other observations given in GUOC report\cite{saf3}. If there are four, five
and six such beads, these beads will closely occupy the corners of a square, pentagon
and hexagon respective, leaving the center unoccupied, since the these beads
tends to occupy as large area of membrane as possible. Since along the waist
of a axisymmetrical dumbbell shape vesicle, the energy density is in maximum and equally
distributed in the surface; the beads can thus be distributed freely with varying
separation distance. These facts are again evident from the experimental
observations\cite{saf3}. 

In summary, our approach gives a more consistent mechanism of physisorption than the others\cite{saf3}: the maximum reduce of 
free energy can induce attraction of colloidal beads on the appreciated position of the membrane. 
Undoubtedly, if the beads are not on the membrane, there is no such attraction, which is confirmed with the experimental 
observation too\cite{saf3}. This may have significance in cell biology. In fact, 
in biophysics, we have known that the the absorption of protein and ion channels mainly in the concave regions in the cell biomembrane 
such as red blood cell\cite{new}, but these phenomena have not be sufficiently studied before.  
 
We are indebted to Drs. Zhou Haijun and Zhou Xin 
for their enlightening discussions. This
subject is supported by National Natural Science Foundation of China.



\begin{figure}
\psfig{file=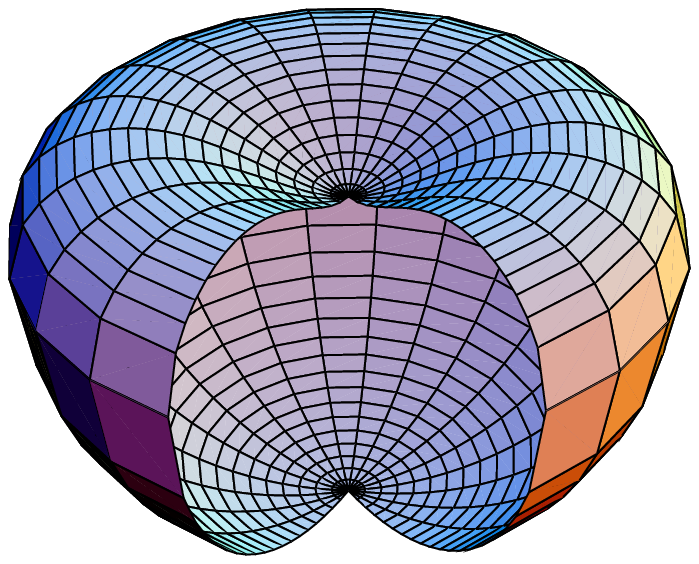,height=80mm,angle=0}
\caption{
A concave region ( only the upper half part is relevant ) fitted those given by experimental ones in
Figs. 1, 3, 5 of GUOC's paper[3]. The parameterization is $ sin\psi(r)=r+2.1 r\ln r $ with $ 10 \mu m $ as the length unit.
}
\label{FIG.1}
\end{figure}

\newpage

\begin{figure}
\psfig{file=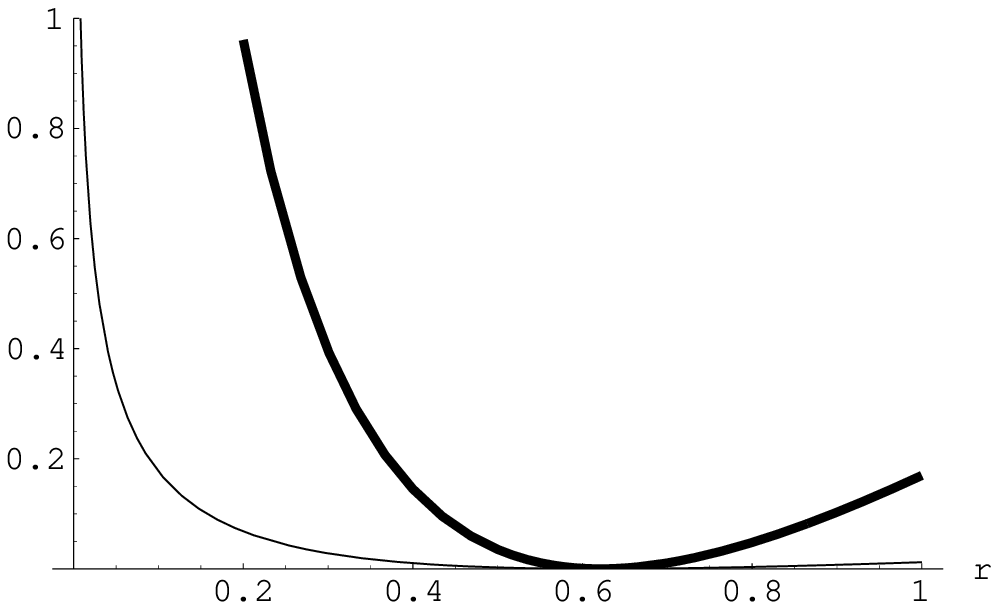,height=80mm,angle=0}
\caption{ The dependence of energy per area and the trapping energy with $r$ in a concave region 
$\sin\psi=r+2.1 r \ln r, r\in (0,10\mu m)$. 
Thin line shows the elastic energy per area using unit $ 5000 k_B T(\mu m)^{-2} $. 
Bold line shows the elastic energy in contact area $\sigma=\pi (0.3)^2(\mu m)^2$ 
(trapping energy) using length unit of $ 10 (\mu m)$ in r axis. }
\label{FIG.2}
\end{figure}


\begin{references}
\bibitem[**]{l}  e-mail address: {\tt liuqh@itp.ac.cn}

\bibitem{saf1}  J. O. R\"{a}dler, I. Koltover, and C. R. Safinya, \
Science {\bf 275}, 810(1997).

\bibitem{saf2}  T.Salditt , I. Koltover, J. O. R\"{a}dler and C. R.
Safinya, \ Phys.\ Rev.\ Lett. {\bf 79}, 2582(1997).

\bibitem{saf3}  I. Koltover, J. O. R\"{a}dler, C. R. Safinya, \ Phys.\
Rev.\ Lett. {\bf 82}, 1991(1999).

\bibitem{sci}  M. S. Spector and J. M. Schnur, \ Scinece {\bf 275},
791(1997).

\bibitem{exp1}  L. Golubovi\'{c} and M. Golubovi\'{c}, \ Phys.\ Rev.\ Lett. 
{\bf 80}, 4341(1998).

\bibitem{exp2}  C.S.O'Hern and T. C. Lubensky, \ Phys.\ Rev.\ Lett. {\bf 80}, 4345(1998).

\bibitem{dan}  N. Dan, Biophys. J. {\bf 71}, 1267 (1996).

\bibitem{oy}  Ou-Yang Zhong-Can and W. Helfrich, Phys.\ Rev.\ Lett.\ E {\bf 59}, 2486 (1987); Phys. Rev. A, {\bf 39}, 5280 (1989).

\bibitem{hel}  W. Helfrich, Z.Naturforsch {\bf 28c}, 693 (1973); H. J.
Deuling and W. Helfrich, Biophys. J. {\bf 16}, 861 (1976); J. Phys. France 
{\bf 37}, 1335 (1976).

\bibitem{oy93}  H. Naito, M.Okuda and Ou-Yang Zhong-Can, Phys.\ Rev.\ E {\bf 48}, 2304 (1993); Phys.\ Rev.\ E {\bf 54}, 2816 (1996).

\bibitem{serf}  U. Serfert, Phys. Rev. {\bf 44}, 1182(1991), Adv. Phys. {\bf 46}, 13(1997), and U. Seifeit and R. Lipowsky, 
in Handbook of Biological
Physics (Vol. I) edited by R. Lipowsky and E. Sackman (Elsevier Science,
Amsterdam, 1995)P.403-P.463.

\bibitem{liu}  Q. H. Liu, Z. Haijun, J.X. Liu and Ou-Yang Zhong-Can
submitted to Phys.\ Rev.\ E.

\bibitem{new}  E. Sackman, in Handbook of Biological
Physics (Vol. I) edited by R. Lipowsky and E. Sackman (Elsevier Science,
Amsterdam, 1995) P.1-P.63. 

\end{references}
\end{document}